# A Coin-Tossing Conundrum

by James Stein (California State University, Long Beach)

and Leonard M. Wapner (El Camino College)

**Abstract**

It is shown that an equiprobability hypothesis leads to a scenario in which it is possible to predict the outcome of a single toss of a fair coin with a success probability greater than 50%. We discuss whether this hypothesis might be independent of the usual hypotheses governing probability, as well as whether this hypothesis might be assumed as a result of the Principle of Indifference. Also discussed are ways to implement or circumvent the hypothesis.

## Introduction

Some mathematical results are surprising, such as the Birthday Problem. Some mathematical results are genuinely startling, such as the Banach-Tarski Theorem or Smale's proof that a sphere can be turned inside out. The Banach-Tarski Theorem is startling because an apparently self-evident hypothesis, the Axiom of Choice, leads to the conclusion that it is possible to double the volume of a sphere through decomposition and rigid motions. In this paper, we discuss an equiprobability hypothesis that appears to lead to the conclusion that it is possible to predict the outcome of a single toss of a fair coin with a success probability greater than 50%.

Diaconis, Holmes, and Montgomery ([2]) have shown that there are dynamical biases in the physical process of flipping a coin. Their approach relies on the physics of the problem. Ours depends on hypotheses and deductions.

## Section 1 – Blackwell's Bet

You are faced with a choice between two identical envelopes, each one containing a different sum of money. Your goal is to choose the one with the larger sum of money, and so you randomly select one of the envelopes and count how much money is in it. You are then allowed to switch your choice for the contents of the other envelope. Should you?

This problem, known as Blackwell's Bet ([3]), appears to be a 50-50 choice. The name for this problem originates from a result of Blackwell ([1]), which can be used to improve the probability of guessing correctly. Select an independent random number from the positive reals to be used as a pointer. If the pointer is greater than the amount of money in the envelope already opened, switch. Otherwise, stick with your original choice.

The mathematics is straightforward. Let G denote the greater amount and L the lesser amount. Let p be the probability that the pointer is less than L, and let q be the probability that the pointer is greater than G. Assume that the two envelopes are equally likely to be selected initially. The probability that L was selected and the pointer is greater than L, indicating that you should correctly switch envelopes, is ½ (1-p). The probability that G was selected and the pointer is less than G, indicating that you should correctly not switch envelopes, is ½ (1-q). The sum of these two probabilities is ½(2-p-q) = 1 - ½(p+q). This is the probability of a successful guess, and is greater than ½ assuming any pointer distribution for which the pointer has a positive probability of falling in any open interval.

There is a scheme by which a coin flip can be integrated with the Blackwell mechanism in such a way that there is a probability of greater than 50% of correctly guessing whether the coin landed heads or tails. Agree that the two envelopes are to be labeled "heads" and "tails", and that the greater amount of money is placed in the envelope corresponding to the outcome of the toss. This results in what might be called a "postdiction" – there is a greater than 50% chance of correctly guessing how the coin actually landed by opening an envelope and selecting an independent pointer to guess which envelope - "heads" or "tails" - contains the greater amount.

This raises an interesting question – can one construct a Blackwell's Bet type mechanism which can predict the result of a yet-to-be-flipped coin with probability greater than 50%?

## Section 2 – The Random Railroad

The Random Railroad is a recasting of the idea of a random walk. The Random Railroad has stations 1 unit apart (although the distance specification is traditional rather than essential to the argument) numbered using successive integers increasing from west to east. Each time the train stops at a station, the engineer flips a coin; if the toss outcome is heads the train proceeds one station to the east, and if it is tails the train proceeds one station to the west.

A passenger goes to sleep on the train. When he wakes up, the train is stopped at a station, and he hears the conductor announce that the next stop is station 4 – but there is no indication of his present location. We will assume that the passenger can discern which way is east. Under the assumption that he is at station 3 or station 5 with equal probability, he can use an independent pointer to guess the direction in which the train is departing. If the pointer is to the west of his current location, he guesses that the train is departing to the west. If it is to the east of his current location, he guesses that the train is departing to the east. Successfully guessing the direction of travel is equivalent to successfully guessing the outcome of the coin toss.

The equal probability hypothesis described in the previous paragraph is central to the argument, and we shall outline some of the possible justifications for this at the end of this section.

The proof that the probability of a successful guess is greater than 50% is essentially identical to the proof of the same statement with regard to Blackwell's Bet. Using the station numbers above without loss of generality, let p be the probability that the pointer is less than 3 and q the probability that the pointer is greater than 5. The probability that he is at station 3 and guesses correctly that he will depart to the east is ½(1-p); similarly the probability that he is at station 5 and guesses correctly that he will depart to the west is ½(1-q), and so again the probability of a successful guess is 1- ½(p+q). If we let r=1-(p+q), then r is the probability that the pointer will fall in the gap between the two stations east and west of the destination. Then p+q = 1-r, and so 1 – ½(p+q) = 1 – ½(1-r) = ½ (1+ r). So the probability of a successful guess can be regarded as the sum of ½ and half the probability that the pointer falls in the gap between the two possible stations of origin.

There is, however, a subtlety in this particular problem that does not appear in the two envelope version of Blackwell's Bet. In Blackwell's Bet, there is no problem comparing the pointer with the money in the envelope, because doing so does not totally reveal which envelope has the greater amount, even though the Blackwell technique of using the pointer does improve the probability of guessing which envelope holds the greater amount. However, in this Random Railroad problem one is not allowed to make a numerical comparison between pointer and station number, because that would presuppose knowledge of the station number. Knowing the station numbers of both current location and destination obviously enables knowledge of the direction of approach to the destination with complete accuracy. Consequently, the passenger must only be able to tell whether the pointer is east or west of his current location, without being able to deduce the station number from it.

From a physical standpoint, this is not difficult. The passenger can use any independent pointer that is east or west of his current location – a flash of light, a bird, anything. There are ways of simulating this without requiring the passenger to make numerical comparisons. We shall postpone the discussion of this until the next section.

Notice that it is the very ambiguity of the passenger's current location – NOT knowing whether he is at station 3 (west of the destination) or station 5 (east of the destination) – that enables him to improve the probability of a successful guess to more than 50%. If the passenger simply knows he is at station 5, and does not know the destination, the pointer does not enable him to guess the direction of travel more than 50% of the time, as he would be guessing the probability that a randomly selected number would be greater or less than 5.

The example presented in this section is also a postdiction of the result of flipping a coin. In order to have an actual prediction of the result of flipping a coin, we must guess whether the coin lands heads or tails before it is flipped.

**The Equal Probability Hypothesis**

The hypothesis that the train starts from station 3 or station 5 with equal probability is central to the proof – and to subsequent discussions. We see several possibilities regarding this hypothesis.

(1) It is simply a hypothesis that can be made about the system we are examining, much as the Axiom of Choice is a hypothesis made prior to the proof of the Banach-Tarski Theorem. The

Axiom of Choice has been shown to be independent of many systems of formal logic; perhaps this hypothesis is similarly independent as well.

(2) It is a consequence of the Principle of Indifference, which is a principle similar to Occam's Razor. The Principle of Indifference states that in the absence of any information about the probability of different outcomes, each outcome should be assumed to have an equal probability.

(3) There are systems in which this hypothesis can be shown to be valid. Example 1, given at the end of this paper, appears to demonstrate this.

(4) There are systems in which this hypothesis may not be needed. .

## Section 3 - Next Stop: Willoughby

We now confront the problem of deciding whether the pointer is east or west of our current location without making any sort of numerical comparison. Assume that, from the standpoint of the passenger, the stations are identified by name rather than number. The passenger has an alphabetical list of station names, and a computer has the list of station names in west-to-east order; that order not being known to the passenger.

While the train is stopped, the passenger looks at the name of the station – Hooterville. The engineer flips the coin, and the conductor informs the passengers that the next stop is Willoughby. Looking at the list of stations, the passenger types one of them, such as Clarksville, into the computer. This choice is made independently and randomly, and requires only a pointer distribution which assigns a positive value to each station.

All the conditions needed for the computation in Section 2 to be correct have been fulfilled. Once again, the passenger assumes his departure station (Hooterville) is east or west of his destination (Willoughby) with equal probability.  However, he knows the destination (analogous to station 4 in the preceding example), and is able to choose an independent pointer which enables him to guess correctly with probability greater than 50% whether the train is headed east or west.  There are two critical elements – equiprobability of current location relative to the destination, and ability to choose an independent pointer which can be assessed as either east or west of the current location – and both are satisfied here.

**Section 4 - Coin Toss Outcome Prediction: The Second Passenger**

While the train is at the current station, but **before** the engineer flips the coin to decide the direction of travel, another passenger sits down beside the passenger in the previous section.  He, too, types the name of one of the cities on the alphabetical station list into the computer, and obtains a pointer – **but he does so before the engineer flips the coin.**  The second passenger can therefore make a prediction about the direction the train will travel – that is, whether the toss outcome will be heads or tails – before the coin is tossed.  But how good is this prediction?

Before the train starts to move, the conductor announces that the destination is Willoughby. The first passenger starts to type a name into the computer, but is interrupted by the second passenger, who says, "Save yourself some trouble – use my pointer."  So both passengers will make the same prediction, as both are simply comparing the independent pointer's location with the current station. Since the first passenger will be right more than 50% of the time, so will the second passenger.

One might argue that there are two possible destinations for the train – Willoughby and whatever station is on the other side of Hooterville.  What if the train went to that other station instead? From the standpoint of both passengers, the name of the destination does not matter, as the train is equally likely to be on either side of the destination – whatever its name may be. No matter what the destination, the two critical elements – from the standpoint of both passengers – are fulfilled: the current location could be on either side of the destination with equal probability, and an independent pointer can be selected which can be discerned to be east or west of the current location.

There is, of course, another way to look at this scenario. The train is at Hooterville, and one possible destination is to the east, the other possible destination is to the west. Therefore, the train is certainly to the west of the eastern possible destination, and to the east of the western possible destination. That would be the point of view of an omniscient onlooker. But the passengers have a different perspective; they have no way to know whether the train is east or west of its destination, whatever that destination may be. Similarly, in Blackwell's Bet, an onlooker with X-ray vision can tell how much money is in each envelope, but the person confronted with the choice of envelopes cannot. The best he can do is improve his probability of guessing correctly, and the passengers are in a similar situation.

## Section 5 – An Example

Although the presentation has assumed that the Random Railroad consists of an east-west track, making this assumption creates certain difficulties. If the track is finite and the train is at an end station, there is only one direction it can go, and so the direction is not determined by the coin toss. If the track is infinite, the assumption that the two possible stations of origin are equiprobable cannot be made for all destinations, as there is no uniform probability distribution on the integers.

We now outline an approach to overcoming the equiprobability problem.

**Example 1**

Assume stations 0 through N are equally spaced clockwise around a circular track. For $1 \leq k \leq N$, let $p_k$ be the probability that the pointer lies on the circular track between station k-1 and station k. Let $p_0$ be the probability that the pointer lies on the circular track between station N and station 0. Although we will use numerical stations to make computations easier to understand, the passenger(s) will always be required to identify stations by name. The coin toss outcome determines whether the train moves clockwise or counterclockwise.

To use the pointer, assume that the passenger is at some station, and has chosen a station from the alphabetical list of stations as described in Section 3. The computer generates a pointer as in Section 3. The computer then chooses a station roughly diametrically opposite the passenger's station, which we will refer to as the reference station RS. The RS and the passenger's station divide the circle into two arcs. The passenger will guess that the train will proceed towards the RS along the arc in which the pointer lies. In the diagram below, the passenger guesses the train will proceed CLOCKWISE in the direction of the arrow. In other words, the passenger is guessing that the destination is the next stop CLOCKWISE.

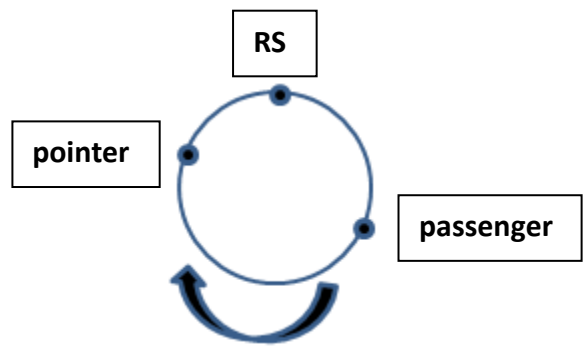

We assume that the train has an equal probability of being at any station. This would be the consequence of random movement of the train. Observe that this arrangement completely disposes of the problem of equiprobability of station of origin, as the train has the same probability of being at any station. We proceed to calculate the probability of the passenger making a correct guess if the destination of the train is station k. We will assume that the RS is station 0; the proof will show that the computation is independent of the location of the RS, as long as it lies outside the minor arc between the two possible stations of origin. We will assume that the station that precedes station k is station k-1 and the station that follows it is station k+1, where 3< k+1<N (this is assured by the choice of the RS).

Consider the following diagram, in which the stations are dots with accompanying station numbers.

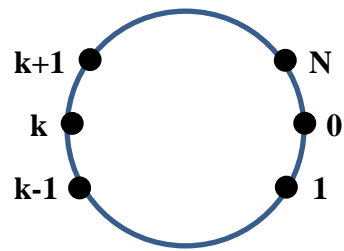

Assuming equiprobability of arrival at station k, the probability of the passenger starting from station k-1 is ½. If the passenger is at station k-1, the probability that the pointer points in the correct direction is

$$p_k + p_{k+1} + \ldots + p_N + p_0$$

So, the probability that the passenger is at station k-1, the destination is station k, and the pointer points in the correct direction (so that the passenger guesses correctly) is

$$½ (p_k + p_{k+1} + \ldots + p_N + p_0)$$

Similarly, the probability of the passenger starting from station k+1 is ½. If the passenger is at station k+1, the probability that the pointer points in the correct direction is

$$p_{k+1} + p_k + \ldots + p_1$$

So, the probability that the passenger is at station k+1, the destination is station k, and the pointer points in the correct direction is

$$\tfrac{1}{2}\,(p_{k+1} + p_k + \ldots + p_1)$$

Therefore, the probability that the destination is station k and the pointer points in the correct direction is

$$\tfrac{1}{2}\,(p_k + p_{k+1} + \ldots + p_N + p_0) + \tfrac{1}{2}\,(p_{k+1} + p_k + \ldots + p_1)$$

$$= \tfrac{1}{2}\,(1 + p_k + p_{k+1})$$

as each term from $p_0$ through $p_N$ appears at least once in the sum, and both $p_k$ and $p_{k+1}$ appear twice.

Consequently, the probability that once the passenger knows the destination is station k, the probability that he will correctly guess the direction of approach is greater than ½. Using the reasoning presented in Section 4, the passenger can guess the direction of approach to the destination with probability greater than ½ even before the destination is decided.

We can actually be a little more precise. The probability that the passenger will correctly guess the direction from a randomly selected (with uniform probability) station can easily be computed as the average of the above quantities. Since there are N+1 stations, the average is

$$\frac{1}{N+1} \sum_{k=0}^{N} \frac{1}{2}(1 + p_k + p_{k+1})$$

$$= \frac{1}{N+1} \left(\frac{N+1}{2} + \sum_{k=0}^{N} \frac{1}{2}(p_k + p_{k+1})\right)$$

$$= \frac{1}{N+1}\left(\frac{N+1}{2} + 1\right) = \frac{1}{2} + \frac{1}{N+1}$$

since the sum from 0 to N of all the $p_k$ (or all the $p_{k+1}$) is 1.

## Section 6 – Open Questions

(1) Is it possible to weaken the equiprobability hypothesis and still derive the conclusion that the passengers can correctly guess the outcome of the coin toss? We believe that this can be done, from continuity considerations alone.

(2) Example 1 was constructed to try to exhibit a configuration in which the equiprobability hypothesis is automatically satisfied. Are there other such configurations? Can we simply mandate the existence of such configurations without the need to construct them, and if so, what does this say about the nature of the equiprobability hypothesis?

One possibility might be to consider a finite track consisting of N stations, with the end stations being viewed as reflecting barriers for a Markov chain. It is easy to see that the steady-state solution is that the two end stations have a probability of 1/(2N-2) of being the train's destination, and the remaining N-2 stations each have a probability of 1/(N-1) of being the train's destination. Therefore, if we simply require the passenger to wake up only if the train is at least 3 stations from an end of the line, the equiprobability condition is fulfilled.

(3) Is it possible to modify or restrict the choice of pointer in such a way to improve the probability of successfully guessing the outcome of the toss?

(4) Can the methods of this paper be used to improve the prediction of the outcome of a toss of a biased coin (one with heads probability different from ½)?

**Conclusion**

One of the reasons for initially posting this on arXiv.org is to solicit help in unraveling what appears to the authors to be an extremely puzzling situation. There is certainly the possibility that we have made some error in the argument, in which case we hope it will be corrected. If the argument appears to be correct, we believe others will be in a better position than ourselves to assess the validity of Example 1 and precisely what role the equiprobability hypothesis plays.

James Stein

Department of Mathematics and Statistics

California State University, Long Beach

Long Beach, CA

james.stein@csulb.edu

Leonard M. Wapner

Division of Mathematical Sciences

El Camino College

Torrance, CA

lwapner@elcamino.edu